\documentclass[aps,prb,showpacs,twocolumn,floats]{revtex4}
\usepackage{graphicx,psfig}
\usepackage{dcolumn}
\usepackage{bm}

\input{tcilatex}

\begin{document}

\title{Interference effect in the Landau-Zener tunneling of the
antiferromagnetically coupled dimer of single-molecule magnets}
\author{D. A. Garanin}
\affiliation{ Institut f\"ur Physik,
Johannes-Gutenberg-Universit\"at,
 D-55099 Mainz, Germany}
\date{\today}

\begin{abstract}
Two antiferromagnetically coupled tunneling systems is a minimal model
exhibiting the effect of quantum-mechanical phase in the Landau-Zener
effect. It is shown that the averaged staying probability oscillates vs
resonance shift between the two particles, as well as vs sweeping rate. Such
a resonance shift can be produced in Mn$_{4}$ dimers by the gradient of the
magnetic field.
\end{abstract}

\maketitle


Tunneling at an avoided level crossing, or the Landau-Zener (LZ) effect \cite
{lan32,zen32} is a quantum phenomenon that was much studied in physics of
atomic and molecular collisions. Recently an experimental technique using
the LZ effect was applied to single-molecule magnets to extract their
tunneling level splitting $\Delta $.\cite{werses99,weretal00epl}

In spite of the quantum nature of the LZ effect, its basic form can be
described classically by a Landau-Lifshitz equation for a magnetic moment in
a time-dependent field.\cite{chugar02,gar03prb,garsch04prb} However
different kinds of interactions between tunneling magnetic molecules in a
crystal make the LZ effect much more complicated. If the interactions are
treated within the mean-field approximation (MFA), then the LZ tunneling can
still be described by a nonlinear Schr\"{o}dinger equation or, equivalently,
by a classical nonlinear Landau-Lifshitz equation. In general, one is left
with a tremendous problem of solving a full Schr\"{o}dinger equation for an $%
N$-particle system.

The question of how good is the MFA for the LZ effect with interaction was
studied in detail for the idealized ``spin bag'' model of $N$ tunneling
particles interacting each with each with the same coupling strength $J$.
\cite{gar03prb,garsch04prb} This model can be mapped onto the problem of the
giant spin $S=N/2,$ so that the MFA limit $N\rightarrow \infty $ corresponds
to the classical limit $S\rightarrow \infty .$ In Ref. %
\onlinecite{garsch04prb} it was shown that if the classical trajectory is
smooth, the MFA yields qualitatively correct results, and quantum
corrections can be calculated with the help of the cumulant expansion. In
the case of complicated classical motion the MFA becomes unreliable.

For the spin-bag model, both the MFA and the full quantum-mechanical
solutions show that the ferromagnetic coupling suppresses LZ transitions
(i.e., increases the LZ staying probability $P$),\ whereas the
antiferromagnetic (AF) coupling increases transitions. This is in accord
with physical expectations based on the time dependence of the total field
on a magnetic molecule, the sum of the external sweep field and the
molecular field from the neighboring molecules. If one of the molecules
tunnels, then for the ferromagnetic coupling the jump of the total field is
positive, so that the neigboring molecules are being brought past the
resonance and lose their chance to tunnel. For the AF-coupling, the jump of
the total field is negative, so that the neigboring molecules recieve an
additional chance to tunnel and thus $P$ decreases.

A more realistic model for the LZ effect with interaction should incorporate
both distance-dependent couplings and individual resonance fields for
magnetic molecules. Recently obtained solution for this model in the
fast-sweep limit \cite{garsch03prl} showed an oscillating dependence of the
averaged one-particle staying probability $P_{\mathrm{avr}}$ of the system
on the resonance shifts between the molecules on the $i$th and $j$th lattice
sites$.$ This is an essentially quantum-mechanical effect arising due to the
possibility to reach the same final state by different sequences of
individual tunneling events. The different quantum-mechanical phases
accumulated on different ways lead to the interference in the final state.
This poses a major challenge in the theoretical description of macroscopic
systems since the phase factors are very sensitive to the microscopic
details and dephasing processes should play a big role. Certainly the effect
of interfering tunneling paths cannot be described by the MFA.

\begin{figure}[t]
\unitlength1cm
\begin{picture}(11,6.5)
\centerline{\psfig{file=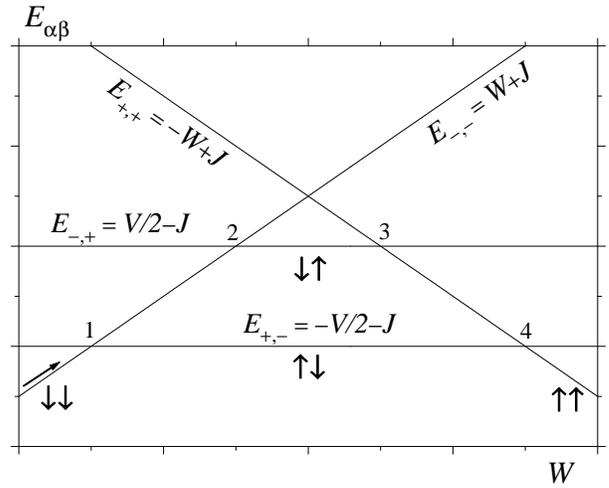,angle=-90,width=10cm}}
\end{picture}
\caption{ \label{Fig-dimer-levels}
Bare energy levels for two antiferromagnetically  coupled tunneling systems for $V<4J$.
}
\end{figure}%
%

The minimal model that exhibits the phase effect in the LZ tunneling is the
model of two antiferromagnetically coupled tunneling systems (see Fig. \ref
{Fig-dimer-levels}) that describes a particular transition in the recently
discovered Mn-4 dimer. \cite{weralihenchr02nature,hiledwalichr03science} We
will use the Hamiltonian
\begin{equation}
\widehat{H}=-\frac{1}{2}\sum_{i=1}^{2}\left\{ \left[ W(t)-V_{i}\right]
\sigma _{iz}+\Delta \sigma _{ix}\right\} +J\sigma _{1z}\sigma _{2z},
\label{Ham}
\end{equation}
where $\mathbf{\sigma }_{i}$ are Pauli matrices, $W(t)=vt\ $is the global
time-linear energy sweep, $\Delta $ is the level splitting, $J$ is the
coupling, and we set
\[
V_{1}=-V/2,\qquad V_{2}=V/2,
\]
so that $V=V_{2}-V_{1}$ is the resonance shift. For $V>0$ particle 1 crosses
the resonance first. The bare energy eigenvalues for this Hamiltonian are \
\begin{eqnarray}
E_{-,-} &=&W(t)+J,\qquad E_{+,+}=-W(t)+J  \nonumber \\
E_{+,-} &=&-V/2-J,\qquad E_{-,+}=V/2-J,  \label{Ealphabeta}
\end{eqnarray}
where $+,-$ means spin 1 up and spin 2 down. There are four first-order
level crossings:
\begin{eqnarray}
E_{-,-} &=&E_{+,-}\quad \Rightarrow \quad W(t)=-V/2-2J=W_{1}  \nonumber \\
E_{-,-} &=&E_{-,+}\quad \Rightarrow \quad W(t)=V/2-2J=W_{2}  \nonumber \\
E_{-,+} &=&E_{+,+}\quad \Rightarrow \quad W(t)=-V/2+2J=W_{3}  \nonumber \\
E_{+,-} &=&E_{+,+}\quad \Rightarrow \quad W(t)=V/2+2J=W_{4}.
\label{Crossings}
\end{eqnarray}
The central crossing, $E_{-,-}=E_{+,+}\quad \Rightarrow \quad W=0$ is the
second-order crossing with the splitting $\sim \Delta ^{2}/J$ that can be
neglected in the case of well-separated resonances.\cite{gar03prb} In this
case one has four independent LZ transitions, each described by a scattering
matrix (see, e.g., Ref. \onlinecite{kay93})
\begin{equation}
M=\left(
\begin{array}{cc}
\sqrt{P} & \sqrt{1-P}e^{-i\phi } \\
-\sqrt{1-P}e^{i\phi } & \sqrt{P}
\end{array}
\right) ,  \label{MDef}
\end{equation}
where
\begin{equation}
P=e^{-\varepsilon },\qquad \varepsilon =\frac{\pi \Delta ^{2}}{2\hbar v}
\label{PLZ}
\end{equation}
is the Landau-Zener staying probability and
\begin{equation}
\phi =\pi /4+\mathrm{Arg}\Gamma \left( 1-i\delta \right) +\delta \left( \ln
\delta -1\right)  \label{MphiDef}
\end{equation}
with $\delta \equiv \varepsilon /(2\pi ).$ Evolution of the wave function
between level crossings reduces to the accumulation of the phase factors $%
\exp \left[ i\Phi _{\alpha \beta }(t)\right] ,$ where the phases
\begin{equation}
\Phi _{\alpha \beta }(t)=-\frac{1}{\hbar }\int^{t}dt^{\prime }E_{\alpha
\beta }(t^{\prime }),\qquad \alpha ,\beta =\pm  \label{Phase}
\end{equation}
can be easily calculated for the linear sweep from Eq. (\ref{Ealphabeta})
and are quadratic in $W.$ The wave function of the system can be written as $%
\Psi (t)=\sum_{\alpha \beta }C_{\alpha \beta }(t)\left| \alpha \beta
\right\rangle .$ Before the first crossing one has both spins down, $%
C_{-,-}=1$ (dropping an irrelevant phase factor) and $C_{\alpha \beta }=0$
otherwise. We use thus defined wave function $C_{\alpha \beta }^{\mathrm{in}%
} $ as the initial condition and we introduce
\begin{equation}
\Delta \Phi _{\alpha \beta }\left( W_{2},W_{1}\right) \equiv \Phi _{\alpha
\beta }\left( W_{2}\right) -\Phi _{\alpha \beta }\left( W_{1}\right) .
\label{DeltaPhialphabeta}
\end{equation}
Then after the fourth crossing one has
\begin{eqnarray}
C_{\alpha ^{\prime \prime }\beta ^{\prime \prime }}^{\mathrm{out}} &=&\left(
\delta _{\alpha ^{\prime \prime },+}M_{\beta ^{\prime \prime }\beta ^{\prime
}}+\delta _{\alpha ^{\prime \prime },-}\delta _{\beta ^{\prime \prime }\beta
^{\prime }}\right)  \nonumber \\
&&\times e^{i\Delta \Phi _{\alpha ^{\prime \prime }\beta ^{\prime }}\left(
W_{4},W_{3}\right) }\left( \delta _{\beta ^{\prime },+}M_{\alpha ^{\prime
\prime }\alpha ^{\prime }}+\delta _{\beta ^{\prime },-}\delta _{\alpha
^{\prime \prime }\alpha ^{\prime }}\right)  \nonumber \\
&&\times e^{i\Delta \Phi _{\alpha ^{\prime }\beta ^{\prime }}\left(
W_{3},W_{2}\right) }\left( \delta _{\alpha ^{\prime },-}M_{\beta ^{\prime
}\beta }+\delta _{\alpha ^{\prime },+}\delta _{\beta ^{\prime }\beta }\right)
\nonumber \\
&&\times e^{i\Delta \Phi _{\alpha ^{\prime }\beta }\left( W_{2},W_{1}\right)
}M_{\alpha ^{\prime }\alpha }C_{\alpha \beta }^{\mathrm{in}},  \label{CoutAF}
\end{eqnarray}
that is the final state. In this formula summation over repeated indices is
implied. The staying probability for the first and second particles are
given by
\begin{eqnarray}
P_{1} &=&1-\left| c_{+,-}\right| ^{2}-\left| c_{+,+}\right| ^{2}  \nonumber
\\
P_{2} &=&1-\left| c_{-,+}\right| ^{2}-\left| c_{+,+}\right| ^{2}.
\label{P1P2Def}
\end{eqnarray}
The average one-particle staying probability and the reduced magnetization
read
\begin{eqnarray}
P_{\mathrm{avr}} &=&\frac{1}{2}\left( P_{1}+P_{2}\right) =1-\frac{1}{2}%
\left| c_{+,-}\right| ^{2}-\frac{1}{2}\left| c_{-,+}\right| ^{2}-\left|
c_{+,+}\right| ^{2}  \nonumber \\
M &=&1-2P_{\mathrm{avr}}.  \label{PAvrDef}
\end{eqnarray}
After some algebra one obtains from Eq. (\ref{CoutAF}) the results
\begin{eqnarray}
&&P_{1}=P^{2}(2-P)  \label{P1P2Res} \\
&&P_{2}=P(2-P)(1-P+P^{2})-2P(1-P)^{2}\cos \left[ \frac{4JV}{\hbar v}\right]
\nonumber
\end{eqnarray}
and
\begin{equation}
P_{\mathrm{avr}}=P\left( 1-\frac{P}{2}\right) (1+P^{2})-P(1-P)^{2}\cos \left[
\frac{4JV}{\hbar v}\right] ,  \label{PAvrRes}
\end{equation}
where the argument of the cos can be rewritten as
\begin{equation}
\frac{4JV}{\hbar v}=\frac{8JV}{\pi \Delta ^{2}}\varepsilon .  \label{ArgCos}
\end{equation}
This is exactly the phase argument in Eq. (8) of Ref. %
\onlinecite{garsch03prl}. The oscillating quantum-phase term in our solution
arises because the state $\left| \downarrow \downarrow \right\rangle $ can
be reached in two different ways: (i) Spin 1 flips first and spin 2 flips
second (crossings at $W=W_{1}$ and $W_{4});$ (ii) Spin 2 flips first and
spin 1 flips second (crossings at $W=W_{2}$ \ and $W_{3}).$ The amplitudes
of these two processes add up, and the accumulated phase difference leads to
oscillations.

The interference effect in a system of two tunneling particles takes place
for the AF coupling only. For the ferromagnetic coupling, the two horizontal
lines corresponding to $\left| \uparrow \downarrow \right\rangle $ and $%
\left| \downarrow \uparrow \right\rangle $ go above the (inactive) cental
crossing, cf. Fig. \ref{Fig-dimer-levels}. As a result, there are only
transitions at crossings at $W_{1}$ and $W_{2}$ [see Eqs. (\ref{Crossings})]
but no transitions at $W_{3}$ and $W_{4}.$ Instead of Eq. (\ref{CoutAF}) one
has
\begin{equation}
c_{\alpha ^{\prime }\beta ^{\prime }}^{\mathrm{out}}=\left( \delta _{\alpha
^{\prime },-}M_{\beta ^{\prime }\beta }+\delta _{\alpha ^{\prime },+}\delta
_{\beta ^{\prime }\beta }\right) e^{i\Delta \Phi _{\alpha ^{\prime }\beta
}\left( W_{2},W_{1}\right) }M_{\alpha ^{\prime }\alpha }c_{\alpha \beta }^{%
\mathrm{in}},  \label{CoutF}
\end{equation}
and the results for the probabilities are
\begin{eqnarray}
P_{1} &=&P,\qquad P_{2}=1-P+P^{2}  \nonumber \\
P_{\mathrm{avr}} &=&(1+P^{2})/2.  \label{P12AvrF}
\end{eqnarray}
Note that $P_{\mathrm{avr}}$ coincides with Eq. (20) of Ref. %
\onlinecite{gar03prb} for $N=2$ and is independent on the resonance shift.
In this model $P_{\mathrm{avr}}\geq 1/2$ because tunneling of both particles
is impossible, $C_{+,+}=0.$ The case of a strong resonance shift corresponds
to $V>4|J|.$ In this case the coupling plays no role, and one obtains $%
P_{1}=P_{2}=P.$

Let us now consider the fast-sweep limit, $\varepsilon \ll 1.$ In this case
one can write the expansion of the averaged staying probability in the form
\begin{equation}
P_{\mathrm{avr}}\cong 1-\varepsilon +\varepsilon ^{2}/2+\varepsilon
^{2}I_{0},  \label{PFastSweep}
\end{equation}
where $I_{0}$ describes the deviation from the standard LZ effect, Eq. (\ref
{PLZ}), due to the interaction. \cite{garsch03prl} For not too large
resonance shifts, $V<4|J|,$ one obtains from Eqs. (\ref{PAvrRes}) and (\ref
{P12AvrF})
\begin{equation}
I_{0}=\left\{
\begin{array}{ll}
-1/2-\cos \left( \frac{8JV}{\pi \Delta ^{2}}\varepsilon \right) , & J>0 \\
1/2, & J<0
\end{array}
\right.  \label{I0Res}
\end{equation}
that is equivalent to Eq. (18) of Ref. \onlinecite{garsch03prl}. For large
couplings and resonance shifts, the cos-term in Eq. (\ref{I0Res}) oscillates
fast and averages out. In this case one can conclude that the effects of
antiferro- and ferromagnetic couplings are just the opposite: The former
enhances transitions while the latter suppresses transitions by the same
amount.

Our present model allows, however, to analyze the influence of interactions
in the whole range of sweep rates, and it shows that the effect of the AF
coupling is smaller than that of the ferromagnetic coupling. Dropping the
cos-term in Eq. (\ref{PAvrRes}), one obtains

\begin{equation}
\overline{P}_{\mathrm{avr}}-P=\left\{
\begin{array}{ll}
-P^{2}\left( 1-P\right) ^{2}/2, & J>0 \\
\left( 1-P\right) ^{2}/2, & J<0.
\end{array}
\right.  \label{PDifference}
\end{equation}
For $J>0$ this difference is small everywhere, and its absolute value
attains a maximum at $P=1/2,$ where $\overline{P}_{\mathrm{avr}}-P=-1/32.$
On the other hand, for the ferromagnetic coupling $\overline{P}_{\mathrm{avr}%
}-P$ \ tends to 1/2 in the slow-sweep limit, $\varepsilon \gg 1.$

\begin{figure}
\unitlength1cm
\begin{picture}(11,6.5)
\centerline{\psfig{file=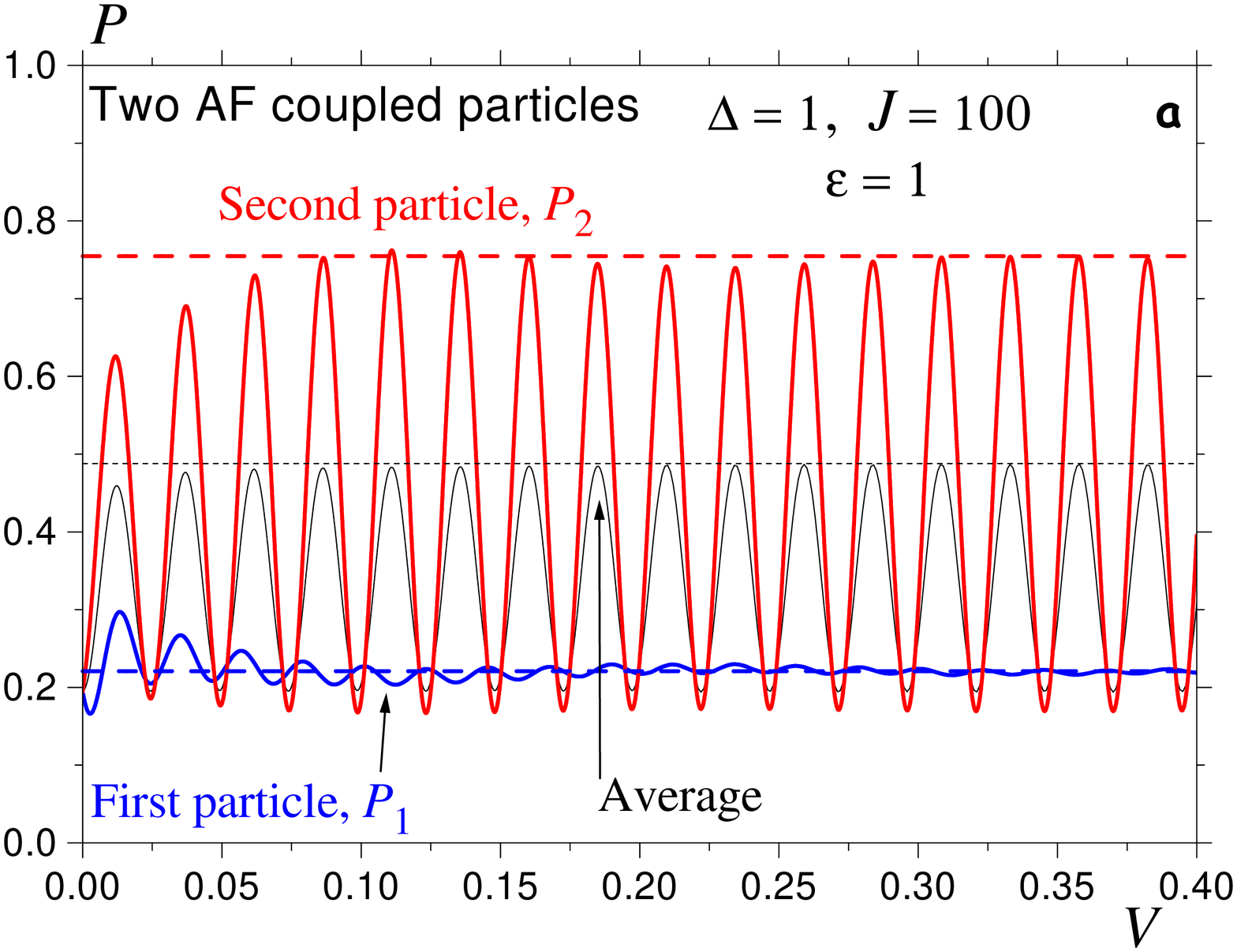,angle=-90,width=9cm}}
\end{picture}
\begin{picture}(11,6)
\centerline{\psfig{file=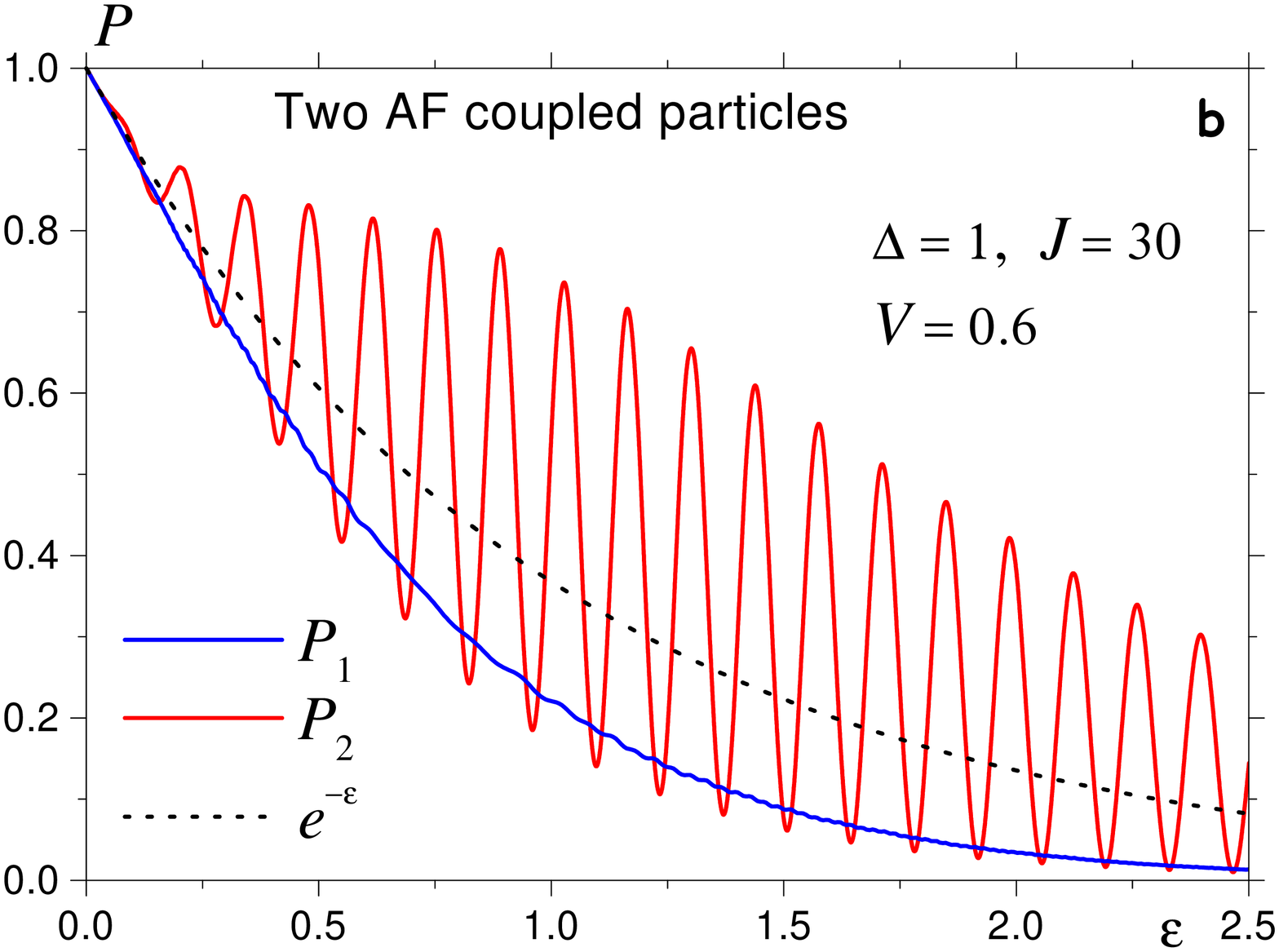,angle=-90,width=9cm}}
\end{picture}
\caption{\label{dimer-PAF} Staying probabilities for a system of two antiferromagnetically coupled particles
$a$ -- vs. resonanse shift $V$. Asymptotic values are shown by horizontal dased lines.
$b$ -- vs sweep parameter $\varepsilon$.
}
\end{figure}%
%

Numerical solution of the LZ problem for two AF-coupled tunneling systems is
shown in Fig. \ref{dimer-PAF}. One can see from Fig. \ref{dimer-PAF}$a$ that
for $J\gg \Delta $ the staying probabilities $P_{2}$ and $P_{\mathrm{avr}}$
begin to oscillate starting from the values of the resonance shift $V$ that
satisfy $V\ll \Delta .$ The numerical data for $P_{\mathrm{avr}}$ \ is well
described by Eq. (\ref{PAvrRes}) starting from $V\approx 0.1.$ On the other
hand, for $P_{1}$ and $P_{2}$ the condition of well-separated resonances is
more restrictive and it requires somewhat greater values of $V$ to validate
Eqs. (\ref{P1P2Res}). Fig. \ref{dimer-PAF}$b$ shows oscillations of $P_{2}$
as a function of the sweep parameter $\varepsilon ,$ as well as a faster
decay of $P_{1}$ in comparison to the standard LZ effect.

Let us now discuss the application of the present results to Mn$_{4}$
dimers. The coupling between the two Mn$_{4}$ monomers with $S=9/2$ was
shown to have the form of the isotropic exchange, \cite{tirwerfogalichr03prl}
$\widehat{H}_{\mathrm{ex}}=J_{\mathrm{ex}}\mathbf{S}_{1}\cdot \mathbf{S}_{2}$
with $J_{\mathrm{ex}}\simeq 0.1$ K for the mostly studied compound.\cite
{weralihenchr02nature,hiledwalichr03science} Density-functional theory
calculations\cite{parketal03prb} yield somewhat larger values of $J_{\mathrm{%
ex}}.$ The uniaxial anisotropy $D\simeq 0.75$ K creates a barrier for spin
tunneling. The level splitting between the ground states $\left| \pm
9/2\right\rangle $ in the Mn$_{4}$ monomer is $\Delta \simeq $ 2$\times
10^{-7}$ K (Ref. \onlinecite{werbhaboschrhen03prbrc}) and it should be of
the same order of magnitude in the Mn$_{4}$ dimer. The ground-state
tunneling in a Mn$_{4}$ dimer can be described by a pseudospin Hamiltonian
of Eq. (\ref{Ham}) with $J=S^{2}J_{\mathrm{ex}}\simeq 2$ K. The period of
oscillations on the resonance shift $V$ follows from Eqs. (\ref{PAvrRes})
and (\ref{ArgCos}) and is given by $V_{\mathrm{Period}}=\left( \pi
^{2}/4\right) \Delta ^{2}/(J\varepsilon ).$ For $\varepsilon =1$ one obtains
$V_{\mathrm{Period}}\simeq 5\times 10^{-14}$ K that corresponds to the
difference of longitudinal magnetic fields $\Delta H\simeq 10^{-14}$ T
between the two monomers. With the distance between the monomers $l\simeq 10$
\AA , this amounts to the very small field gradient, $dH/dx\simeq 10^{-3}$
Gauss/cm! This means that small inhomogeneities of the magnetic field, due
to, e.g., dipole-dipole interaction, should average out the quantum
oscillations in the LZ effect. For possible applications in quantum
computing, it is desirable to have a larger tunnel splitting $\Delta ,$ to
achieve a faster performance rate and reduce the influence of decoherence.
This can be achieved by applying a transverse magnetic field. In this case
the period of the quantum oscillations considered here will be much larger,
and their observation will require much robuster field gradients that will
exceed uncontrolled inhomogeneities of the magnetic field.

The author thanks R. Schilling for many stimulating discussions.

\bibliographystyle{prsty}

\end{document}